\documentclass{hep99}
\usepackage{epsf}
\begin{document}
\newcommand{\be}{\begin{equation}}
\newcommand{\ee}{\end{equation}}
\newcommand{\bea}{\begin{eqnarray}}
\newcommand{\eea}{\end{eqnarray}}
\newcommand{\nn}{\nonumber}
\newcommand{\dd}{\displaystyle}

\title{Bounds on new physics from  parity violation  in
atomic cesium$^{\dag}$}

\author{Daniele Dominici$^{1,2}$}
%

\address{$^1$ Dipartimento di Fisica, Universit\`a di Firenze, I-50125
Firenze, Italia\\
$^2$ I.N.F.N., Sezione di Firenze, I-50125 Firenze, Italia\\[3pt]
E-mails: {\tt dominici@fi.infn.it}}

\abstract
{A recent experimental determination of
the weak charge of atomic cesium is used to get 
implications for possible
new physics. 
 The new data  imply  positive upper and lower
bounds on the new physics contribution to the weak charge, $\delta_NQ_W$,
requiring  new physics
 of a type not severely constrained by the high energy precision data.
} 

\maketitle

\fntext{\dag}{
Talk given at the International 
Europhysics Conference on High Energy Physics, Tampere,
Finland, 15-21 July 1999.}

\section{Introduction}
In a recent paper \cite{bennett} a new determination of the weak
charge of atomic cesium has been reported. 
The most precise parity violating (PV) experiment compares the
mixing among $S$ and $P$ states due to neutral weak interactions
to an induced Stark mixing \cite{wood}. The 1.2\% uncertainty
on the weak charge $Q_W$ was dominated by the
theoretical 
calculations on the amount of Stark mixing  and on the
electronic PV matrix elements. In this recent paper \cite{bennett}
the Stark mixing was measured and, incorporating new experimental
data, the uncertainty in the electronic PV matrix elements
was reduced.
The new result
\be
Q_W(^{133}_{55}Cs)=-72.06\pm (0.28)_{\rm expt}\pm (0.34)_{\rm
theor}
\label{newexp}
\ee
represents a considerable improvement with respect to the previous
determination \cite{noecker,blundell}
\be
Q_W(^{133}_{55}Cs)=-71.04\pm (1.58)_{\rm expt}\pm (0.88)_{\rm
theor}
\ee

On the theoretical side, $Q_W$ can be expressed in terms of the
$S$ parameter  \cite{marciano} or the $\epsilon_3$
\cite{altarelliqw}
\be
Q_W=-72.72\pm 0.13-102\epsilon_3^{\rm rad}+\delta_NQ_W
\ee
including hadronic-loop uncertainty. We use here the variables
$\epsilon_i$ (i=1,2,3) of ref. \cite{altarelli}, which include the
radiative corrections, in place of the set of
variables $S$, $T$ and $U$ originally introduced in ref. \cite{peskin}.
In the above
definition of $Q_W$
we have  explicitly included only the Standard Model (SM) contribution
to the radiative corrections. New physics (that is physics beyond the
SM) contributions
to $\epsilon_3$ are represented by the term $\delta_N Q_W$. Also, we
have neglected a correction proportional to $\epsilon_1^{\rm rad}$.
In fact, as well known \cite{marciano}, due to the
particular values of the number of neutrons ($N=78$) and of
protons ($Z=55$) in cesium, the dependence on
$\epsilon_1$ almost cancels out.

From the theoretical expression
we see that $Q_W$ is particularly sensitive to new physics
contributing to the parameter $\epsilon_3$. This kind of new
physics is severely constrained by the high energy experiments.
From a recent analysis \cite{altarelli2}, one has that
the  value of $\epsilon_3$ from the  high
energy data is
\be
\epsilon_3^{\rm expt}=(4.19\pm 1.0)\times 10^{-3}
\ee
To estimate new physics contributions to this parameter one
has to subtract the SM radiative corrections, which, for
$m_{top}=175~GeV$ and for $m_H~(GeV)=100,~300$, are given
respectively by
\be
\matrix{m_H=100~GeV && \epsilon_3^{\rm rad}=5.110\times 10^{-3}\cr
        m_H=300~GeV && \epsilon_3^{\rm rad}=6.115\times
        10^{-3}}
\ee
Therefore new physics contributing to $\epsilon_3$
cannot be larger than a few per mill. Since $\epsilon_3$ appears
in $Q_W$ multiplied by a factor 102, this kind of new physics which
contributes through $\epsilon_3$
cannot contribute to $Q_W$ for more than a few tenth. On the
other side the discrepancy between the SM and the experimental
data is given by (for a light Higgs)
\be
Q_W^{\rm expt}-Q_W^{SM}=1.18\pm 0.46
\ee
where we have added in quadrature the uncertainties. 
Therefore the 95\% CL limits on $\delta_NQ_W$ are
\be
0.28\leq \delta_NQ_W\leq 2.08
\label{bounds}
\ee
For increasing $M_H$ both bounds increase.
These bounds have been used recently \cite{apv,rosner} 
to get implications 
on new physics and will be reviewed here.

\section{Bounds on new physics}

Let us now look at models which, at least in principle, could
give rise to a sizeable modification of $Q_W$. In ref.
\cite{altarelli3} it was pointed out that models involving extra
neutral vector bosons coupled to ordinary fermions can do the
job.
The high energy data at the Z resonance strongly bound the $Z-Z'$ mixing 
\cite{gross}.
 For this
reason we will assume zero mixing  in the following calculations.
In this
case $\delta_NQ_W$ is completely fixed by the $Z^\prime$
parameters: 
\be
\delta_NQ_W=
16 a_e^\prime [(2Z+N)
v_u^\prime+
(Z+2N)  v_d^\prime]
\frac{M_Z^2}{M_{Z^\prime}^2}
\label{deltaQ}
\ee
$a_f^\prime,
v_f^\prime$ are the couplings  $Z^\prime$ to fermions.

We will discuss three classes of models: the left-right (LR) models,
the extra-U(1) models,  and the so-called sequential SM
models  (that is models with fermionic couplings just scaled from those
 of the SM).
\begin{table}[bht]
\caption{Vector and axial-vector coupling constants for the determination of
$\delta_NQ_W$ for the various models considered in the text. 
 The different
extra-U(1) models are parameterized by the angle $\theta_2$, and
in the table $c_2=\cos\theta_2$, $s_2=\sin\theta_2$. This angle
takes a value between $-\pi/2$ and $+\pi/2$.}
\begin{center}
\begin{tabular}{|c|c|}
\hline
Extra-U(1) & LR \\
\hline\hline
 $a_e^\prime=\frac 1 4 s_\theta\left(-\frac 1 3
c_2+\sqrt{\frac 5 3} s_2\right)$&$a_e^\prime=-\frac 1 4
\sqrt{c_{2\theta}}$\\
\hline
$ v_u^\prime=0 $&$
v_u^\prime=\frac{\dd \left(\frac 1 4 - \frac 2 3
s_\theta^2\right)}{\dd \sqrt{c_{2\theta}}}$\\
\hline
 $v_d^\prime=\frac 1 4 s_\theta\left( c_2+\sqrt{\frac 5 3}
s_2\right)$ &$  v_d^\prime=\frac {\dd \left(-\frac 1 4 +
\frac 1 3 s_\theta^2\right)}{\dd \sqrt{c_{2\theta}}}$\\
\hline
\end{tabular}
\end{center}
\end{table}

In the case of the  LR model  we get a contribution
\be
\delta_NQ_W=-\frac{M_Z^2}{M_{Z^\prime}^2}Q_W^{SM}~~~~
\ee
For this model one has a 95\% lower bound on $M_{Z^\prime}$ from
Tevatron \cite{tevatron} given by $M_{Z^\prime}\ge 630~GeV$.

A LR model could explain the data allowing for a mass
of the $Z^\prime$ varying between the intersection from the 95\%
CL bounds $540\le M_{Z^\prime}(GeV)\le 1470$
deriving from eq. (\ref{bounds})
and the lower bound of $630~GeV$ .

In the case of the extra-U(1) models  the CDF experimental lower
bounds for the masses vary according to the values of the
parameter $\theta_2$ which parameterizes different
extra-U(1) models, but in general they are about
$600~ GeV$ at 95 \% CL \cite{tevatron} (see Fig. 1).
 From eq. (\ref{deltaQ})
 we can easily see that  the models with $\theta_2$
in the interval   $-0.66\le\theta_2({\rm rad})\le 0.25$  give
$\delta_NQ_W\le 0$, and therefore they are excluded at the 99\%
CL. In particular the models known in the literature as  $\eta$
(or $A$),
 which corresponds to $\theta_2=0$, and 
  $\psi$
(or  $C$), which corresponds to 
$\theta_2=-0.66$, are excluded.
 
The bounds on $\delta_N Q_W$  at 95 \%
CL  can be translated into  lower and upper bounds on
$M_{Z^\prime}$. The result is given in Fig. 1, where the bounds
are plotted versus $\theta_2$. In looking at this figure one
should also remember that the direct lower bound from Tevatron is
about 600 ~GeV at 95\% CL.
 The $\chi$ (or $C$) model, corresponding to  $\theta_2=0.91$, is still
allowed.

The last possibility we consider is a sequential SM. In this case
we assume that the couplings are the ones of the SM just scaled
by a common factor $a$. Therefore we get
\be
\delta_NQ_W=a^2 \frac{M_Z^2}{M_{Z^\prime}^2}Q_W^{SM}
\ee
We see that no matter what the choice of $a$ is, the sign of the new physics
contribution turns out to be negative. Therefore all this class
of models are excluded at 99\% CL.

Finally we have considered  
 certain models based on extra dimensions which have a tower
of Kaluza-Klein resonances of the $W$ and $Z$ with masses in the $TeV$
range \cite{qui,KK}. These large extra dimensions appear in the 
string theory context
or as a framework to break supersymmetry.
In the more general case with two higgs (one in the bulk and one on
the wall) PV data put a lower limit on the mixing angle of the
KK modes with the SM gauge bosons allowing only the region of
maximal mixing \cite{KKnoi} ($\sin\beta\geq 0.707$ at 95\% CL).

Another interesting possibility one can analyze is that of a four-fermion
contact interaction, which could arise from different theoretical origins.
Also this case has no visible effects at the
$Z$ peak. We will follow  the analysis and the notations of
ref. \cite{langacker}. In this situation it turns out to be
convenient to express the weak charge as
\be
Q_W=-2\left[c_{1u}(2Z+N)+c_{1d}(Z+2N)\right]
\ee
where $c_{1u,d}$ are products of vector and axial-vector
couplings. We will consider models with a contact
interaction given by
\be
{\cal L}=\pm \frac{4\pi}{\Lambda^2}\bar e\Gamma_\mu e\bar q\Gamma^\mu q,~~~~
\Gamma_\mu=\frac 1 2\gamma_\mu(1-\gamma_5)
\ee
This leads to a shift in the couplings given by
\be
c_{1u,d}\to c_{1u,d}+\Delta C,~~~~\Delta C=\mp\frac{\sqrt{2}\pi}{G_F\Lambda^2}
\ee
Since a variation of the couplings induces a variation of $Q_W$
of opposite sign, we see that the choice of the negative sign in
the contact interaction is excluded. In the case of the positive
sign, using  the 95\% CL bounds given in eq. (\ref{bounds}),  we
get
$
12.1\le\Lambda^+(TeV)\le 32.9
$
to be compared with the PDG limit $\Lambda^+(TeV)\ge 3.5~TeV$.

Let us now consider a contact interaction induced by
lepto-quarks. Following again ref. \cite{langacker}, we take the
case of so-called $SU(5)$-inspired leptoquarks, leading to the interaction
\be
{\cal L}=\frac{\eta_L^2}{2M_S^2}\bar e_L\gamma_\mu e_L\bar
u_L\gamma^\mu u_L+\frac{\eta_R^2}{2M_S^2}\bar e_R\gamma_\mu
e_R\bar u_R\gamma^\mu u_R
\ee
From the constraints on $\pi_{e2}/\pi_{\mu 2}$ one expects
$\eta_L\approx 0$ or $\eta_R\approx 0$. Only the coupling
$c_{1u}$ has a shift
\be
c_{1u}\to c_{1u}+\Delta C,~~~~
\Delta C= \mp\frac{\sqrt{2}\eta_{L,R}^2}{8G_F M_S^2}
\ee
It follows that the shift on $Q_W$ is negative for
$\eta_R\not=0$. Therefore only the left coupling is allowed
($\eta_R=0$). In that case we get the bounds (again from eq.
(\ref{bounds}))
$
1.7\le {M_S(TeV)}/{\eta_L}\le 4.5
$.
If one assumes $\eta_L^2\approx 4\pi\alpha$, it follows
$
0.5\le M_S(TeV)\le 1.2
$.

\begin{figure}
\begin{center}
\epsfysize=7truecm
\epsfxsize=7truecm
\centerline{\epsffile{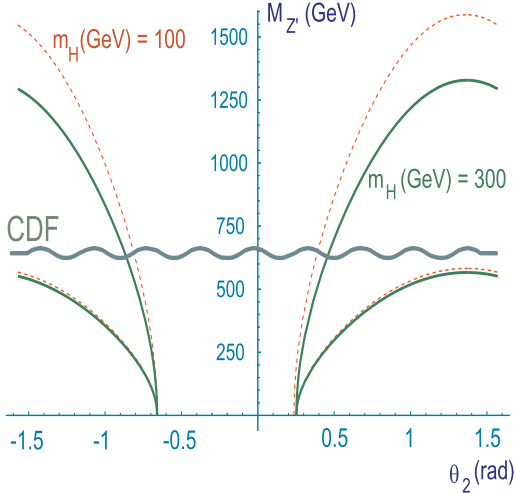}}
\end{center}
\caption{The 95\% CL lower and upper bounds for $M_{Z^\prime}$ for the
extra-U(1) models versus $\theta_2$. The continuous and the
dashed lines correspond to $m_H=100~GeV$ and $m_H=300~GeV$
respectively. CDF lower bound is also shown.}
\end{figure}

\section*{Acknowledgments}
I would like to thank 
R. Casalbuoni, S. De Curtis and  R. Gatto 
for the fruitful and enjoyable collaboration on the topics covered here
and E. Gross for interesting discussions.

\end{document}